\documentclass[conference,compsoc]{IEEEtran}
\IEEEoverridecommandlockouts

\usepackage{cite}
\usepackage{amsmath,amssymb,amsfonts}
\usepackage{graphicx}
\usepackage{textcomp}
\usepackage[T1]{fontenc}
\usepackage{newtxtext,newtxmath}
\usepackage{booktabs,array,url,xcolor,colortbl,pifont,multirow}
\newcolumntype{L}[1]{>{\raggedright\arraybackslash}p{#1}}  
\usepackage{tikz}
\usetikzlibrary{positioning,arrows.meta,fit,calc,shapes.geometric,shapes.misc}
\usepackage[hidelinks]{hyperref}
\definecolor{blu}{HTML}{2A78D6}\definecolor{ink2}{HTML}{52514E}\definecolor{mut}{HTML}{9A9993}\definecolor{warm}{HTML}{EB6834}

\newcommand{\rb}{RB}

\providecommand{\pend}[1]{\textcolor{red}{#1}}
\usepackage[most]{tcolorbox}
\newtcolorbox{rqanswer}{colback=black!6,colframe=black!35,boxrule=0.4pt,arc=2pt,left=3pt,right=3pt,top=1pt,bottom=1pt,before skip=3pt,after skip=3pt}

\definecolor{promptbg}{HTML}{F4F4F1}
\newtcolorbox{promptbox}[1][]{breakable,colback=promptbg,colframe=promptbg,boxrule=0pt,arc=1pt,left=3pt,right=2pt,top=2pt,bottom=2pt,before skip=4pt,after skip=6pt,fontupper=\scriptsize,before upper={\setlength{\topsep}{0pt}\setlength{\partopsep}{0pt}},#1}
\begin{document}

\title{Reinforcement Building: Toward Self-Improving Harnesses for Defensive Software Security Agents}

\author{
    Amine Lbath\textsuperscript{\rm 1,\rm 2},
    Aurelien Delaitre\textsuperscript{\rm 1},
    Vadim Okun\textsuperscript{\rm 1}, \\
    Massih-Reza Amini\textsuperscript{\rm 2}
    \\
    \textsuperscript{\rm 1}National Institute of Standards and Technology,
    \textsuperscript{\rm 2}Université Grenoble Alpes, CNRS
    \\
    \thanks{Disclaimer: Certain trade names and company products are mentioned in the text or identified. In no case does such identification imply recommendation or endorsement by the National Institute of Standards and Technology (NIST), nor that they are necessarily the best available for the purpose.}
}

\maketitle

\begin{abstract}
Large language model (LLM) agents have become capable of discovering and exploiting vulnerabilities in real-world software with limited human intervention, which widens an asymmetry that already favors attackers. Defenders therefore need markedly better repair agents to keep up with the growing number of discovered vulnerabilities. Specialized multi-agent harnesses have been shown to improve repair agents, but they are often crafted for a specific model and do not necessarily transfer to other models. Given the pace at which new models are released, rebuilding a harness by hand for each of them appears impractical.

We present Reinforcement Building (RB), a framework that automatically builds the harness of a defensive agent for a given model. In each round, RB runs the current harness on training vulnerabilities and grades every patch with an execution oracle. An optimizer model then proposes edits to its agents, tools, delegation edges and memory, and a validation gate keeps them only if they strictly improve the harness. Without updating any weights, RB improves the repair success rate of three open-weight models on three held-out benchmarks by a median of 77\% relative to the base harness. On SEC-bench, it outperforms ContraFix, the state-of-the-art harness builder for vulnerability repair, by 21\% relative on both Gemma-4 models. Compared with reinforcement learning with verifiable rewards (RLVR) on the same data, RB needs 7.4 times fewer training episodes to reach the same success rate on Gemma-4-31B. \\
Code and data are available \href{https://reinforcement-building.github.io/}{reinforcement-building.github.io/}.
\end{abstract}


\section{Introduction}
\label{sec:intro}

\begin{figure}[!b]
\centering
\includegraphics[width=\columnwidth]{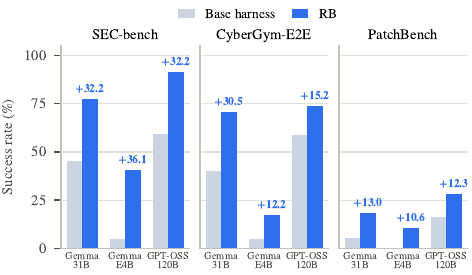}
\caption{Success rate of each model with the base harness and with the harness built for it by \rb{} on the three held-out benchmarks. Labels give the gain in percentage points. Every gain is significant (95\% confidence interval excludes zero).}
\label{fig:gains}
\end{figure}

Large language model (LLM) agents have become capable of complex offensive security tasks. Recent agents have discovered vulnerabilities in real projects, generated proof-of-concept inputs and exploited known and unknown vulnerabilities with limited human intervention~\cite{sok,cybergym,pagent,aixcc}. This widens an asymmetry that already favors attackers. An attacker needs a single overlooked vulnerability, whereas defenders must find and fix vulnerabilities before they are exploited. Defenders therefore need markedly better repair agents to keep up with the growing number of discovered vulnerabilities~\cite{sok}.

The strongest repair systems have placed the model inside a specialized harness that coordinates agents, tools and instructions, and these harnesses have been shown to improve the success rate over a base harness~\cite{patchagent,san2patch,contrafix}. A specialized harness, however, typically compensates for the weaknesses of the model it was designed for, and different models have different weaknesses. Its gains therefore do not necessarily translate to other models. Since new models are being released every few weeks, rebuilding a harness by hand for each of them would be impractical, and defensive teams have had to reuse harnesses designed for other models.

Reinforcement learning with verifiable rewards (RLVR) can instead adapt each new model to the task automatically, but it requires updating the model weights, which is impossible for closed-weight models and needs heavy GPU training for open-weight ones. It also learns from a scalar reward. When the model rarely succeeds, a binary reward is almost always zero, and RLVR has little to learn from. Shaped rewards can give partial credit, for example for a patch that builds or for avoiding a known failure behavior, but they must be designed in advance, they can be gamed, and a reward designed for the failures of one model does not necessarily address those of another. The trace of a failed episode, however, can provide a useful signal on where the agent failed, which could be leveraged to improve the next attempts.

Because weight updates are not possible for every model and are costly to repeat for each new one, recent work has instead started to focus on methods that automatically improve repair harnesses for defensive security tasks. EvoRepair~\cite{evorepair} extracted experiences from past repair trajectories and retrieved them to guide later repairs, and ContraFix~\cite{contrafix} stored validated repair episodes as reusable skills. Both have achieved state-of-the-art results on patching vulnerabilities in SEC-bench~\cite{secbench}. Both, however, focused on adding past experiences or skills to the context of the agent, while leaving its tools, sub-agents and runtime controls unchanged.

\begin{figure*}[t]
    \centering
    \includegraphics[width=\textwidth]{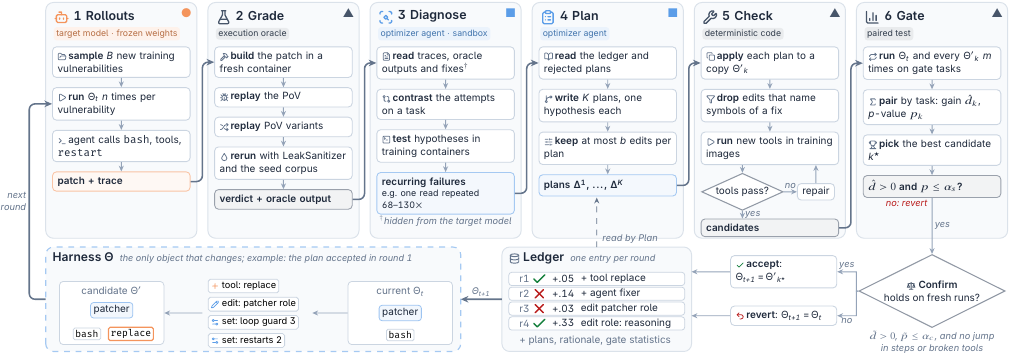}\vspace{-1mm}
    \caption{Overview of one Reinforcement Building round. The tag and the symbol of each step name its actor: the target model (\textcolor[HTML]{F4945C}{\ding{108}}, orange), the optimizer model (\textcolor[HTML]{5B9DE8}{\ding{110}}, blue) and deterministic code (\textcolor[HTML]{475569}{\ding{115}}, grey). Examples are taken from the build for Gemma-4-31B, and the ledger shows the gain $\hat d$ of the selected plan in rounds 1 to 4.}
    \label{fig:overview}
\end{figure*}

We propose Reinforcement Building (\rb{}), a new framework that learns specialized harnesses for given models from execution feedback (Fig.~\ref{fig:overview}). The successful and failed episodes of the model provide signals on which parts of the harness to change, from instructions, to tools, memory and sub-agents, and held-out tasks indicate whether a change helps. \rb{} keeps the iterative structure of RLVR, which guides the build round by round, and replaces the weights with the harness components. In each round, the target model runs the current harness on a batch of training vulnerabilities, attempting each vulnerability multiple times, and an execution oracle grades every patch. An optimizer agent reads the traces of all runs, contrasts failed and successful attempts on the same vulnerability, attributes recurring failures to components of the harness and proposes alternative plans of edits to the harness, such as adding a tool or rewriting an instruction. A statistical gate accepts a plan only if it improves the harness on held-out gate tasks, and a ledger of past plans and their outcomes informs the next round.

We investigate the following research questions:
\begin{itemize}\setlength{\itemsep}{0pt}\setlength{\topsep}{2pt}\setlength{\parsep}{0pt}
\item[] \textbf{RQ1:} How does \rb{} compare with state-of-the-art harness builders for vulnerability repair and with RLVR?
\item[] \textbf{RQ2:} How efficiently does \rb{} learn, and how does the harness evolve over rounds?
\item[] \textbf{RQ3:} What changes does \rb{} make to the behavior of the agent?
\item[] \textbf{RQ4:} Which components of the built harnesses contribute to the gains?
\item[] \textbf{RQ5:} What role do the optimizer and the target model play in the gains of \rb{}?
\item[] \textbf{RQ6:} Does a harness built for one model transfer well to other models?
\end{itemize}

Across three held-out benchmarks and three models (Fig.~\ref{fig:gains}), \rb{} improved the success rate by 26\% to 750\% relative to the base harness (10.6 to 36.1 percentage points). With the same training data, it outperformed the state-of-the-art harness builders for vulnerability repair, ContraFix and EvoRepair, by 24\% and 66\% relative on average across the two Gemma-4 models and the three benchmarks. Compared with RLVR, \rb{} needed at least 7.4 times fewer training episodes to reach the same success rate on SEC-bench. Finally, we observed that a harness built for one model does not necessarily transfer to other models, and that each model gains most from its own harness.

In this paper, we make the following contributions:
\begin{itemize}\setlength{\itemsep}{1pt}\setlength{\topsep}{2pt}
\item We present Reinforcement Building (\rb{}), a framework inspired by RLVR that learns model-optimized harnesses to repair vulnerabilities, including their instructions, tools and sub-agents, from execution feedback and without updating the model weights (Section~\ref{sec:rb}).
\item We evaluate \rb{} on three models and three held-out, execution-grounded benchmarks against RLVR and the state-of-the-art harness builders for vulnerability repair, all trained on the same dataset, and we analyze how each method changes the behavior of the agent, how built harnesses transfer across models and which components produce the gain (Section~\ref{sec:results}).
\item We release the code of \rb{}, the built harnesses, the build ledgers and all graded evaluation results.\footnote{\url{https://github.com/reinforcement-building/reinforcement-building}}
\end{itemize}
All vulnerabilities in this work are public, and all experiments ran in isolated containers.

\section{Background}
\label{sec:background}

\textbf{Vulnerability repair and its evaluation.} Automated vulnerability repair aims to fix a vulnerability in a program without human help and without breaking its functionality. In recent benchmarks, a repair task pairs a real vulnerability with the project at the vulnerable commit, a build script and a proof-of-vulnerability (PoV) that triggers the vulnerable behavior, which mirrors how vulnerabilities reach maintainers in practice. The agent works inside the project's container and submits a patch. An execution oracle accepts the patch if the project builds and the PoV no longer triggers the vulnerability, and some benchmarks also run the tests of the project~\cite{secbench,arvo,patchbench}. The verdict is binary and computed by running code, so it can serve as a verifiable reward.

\textbf{Harnesses.} A harness is everything around a model that turns it into an agent. It contains the instructions of each agent, the tools exposed as function calls, the sub-agents and the edges along which agents delegate, the memory that agents read, and runtime controls such as step limits. The base harness of SEC-bench is a single agent with a shell and a task description~\cite{secbench}. Published repair systems have added localization agents, verification tools, multi-candidate generation and retrieval of past repairs~\cite{patchagent,san2patch,contrafix,evorepair}.

\textbf{Reinforcement learning with verifiable rewards.} RLVR is a post-training technique that improves a model on tasks whose outcome can be checked automatically. It samples multiple episodes per task from the current policy, scores them with a predefined verifiable reward and updates its weights toward the episodes that score above the group mean~\cite{grpo,deepseekr1,swerl}. Low-rank adapters reduce the cost of the update by training small low-rank matrices added to the frozen weights~\cite{lora}. The update needs access to the weights and GPU training for each model, and its learning signal relies solely on the scalar reward.

\textbf{Weight-space programming.} Model editing changes a small set of weights to alter a specific behavior of a model, without gradient training. Locate-then-edit methods first identify the components that drive the behavior, such as hidden units of an MLP layer or a direction in the activations, and then apply a closed-form change to the corresponding weights~\cite{rome,memit,arditi}. Weight-space programming (WSP) has recently delegated this process to an LLM, which reads the weights and activations of a network and writes the edit itself~\cite{wsp}. An edit is cheap and targeted, but it requires access to the weights and a behavior that can be located in them.

\section{Reinforcement Building}
\label{sec:rb}

This section describes how Reinforcement Building works, including its harness representation, its objective and its build loop (Fig.~\ref{fig:overview}).

\textbf{Harness representation.} \rb{} represents a harness $\Theta$ as a structured specification that lists agents with their roles, notes and step limits, tools exposed as function calls, delegation edges, shared memory and an entry agent. It also sets control parameters of the runtime, such as a loop guard that interrupts an agent that repeats the same command, the number of times an agent may start over from the original code, keeping only a note on what it learned, and a call budget for each sub-agent. The runtime executes $\Theta$ inside the task's container. The optimizer changes $\Theta$ only through a fixed set of operations that add, edit or remove agents, tools and notes, grant or revoke edges, and set control parameters, so it can change the structure of the harness as well as its text.

\textbf{Objective.} Let $M$ be the target model with frozen weights, $\mathcal{D}$ a distribution of vulnerabilities and $\mathcal{O}$ the execution oracle. The verifiable reward of a harness $\Theta$ on a task $x$ is $r(\Theta,x)=\mathcal{O}\big(x,y_{M,\Theta}(x)\big)\in\{0,1\}$, where $y_{M,\Theta}(x)$ is the patch that $M$ produces with $\Theta$. \rb{} seeks the harness $\Theta^{\star}$ that maximizes the expected reward:
\begin{equation}
\Theta^{\star} \in \arg\max_{\Theta}\ \mathbb{E}_{x\sim\mathcal{D}}\left[r(\Theta,x)\right].
\label{eq:objective}
\end{equation}
Since $\mathcal{D}$ is unknown, \rb{} uses training vulnerabilities to propose changes to the harness and held-out gate tasks to estimate the expected reward of each candidate.

\textbf{Build loop.} Each round of \rb{} mirrors an RLVR update~\cite{grpo,deepseekr1}, with the harness in place of the weights (Table~\ref{tab:analogy}). The difference lies in the learning signal. RLVR compares the rewards of a group of episodes and moves the weights toward the better ones. \rb{} keeps the reward as the acceptance criterion, but its optimizer also reads each episode, contrasts failed and successful attempts and attributes each failure to components of the harness. A single failed episode therefore carries information even when no episode of the batch succeeds. A round has six steps (Fig.~\ref{fig:overview}).

\begin{table}[!htbp]
\caption{Correspondence between an RLVR update and a round of \rb{}.}
\label{tab:analogy}
\centering\footnotesize
\setlength{\tabcolsep}{3pt}
\begin{tabular}{@{}L{0.27\columnwidth}L{0.31\columnwidth}L{0.36\columnwidth}@{}}
\toprule
Concept & RLVR & \rb{} \\
\midrule
Parameters & model weights & harness $\Theta$ (agents, tools, notes, edges) \\ \arrayrulecolor{black!12}\specialrule{0.3pt}{1.5pt}{1.5pt}\arrayrulecolor{black}
Batch & batch of training tasks per gradient step & batch of $B$ training vulnerabilities per round \\ \arrayrulecolor{black!12}\specialrule{0.3pt}{1.5pt}{1.5pt}\arrayrulecolor{black}
Rollouts & multiple sampled episodes per training task & $n$ episodes per training vulnerability, run with the current harness \\ \arrayrulecolor{black!12}\specialrule{0.3pt}{1.5pt}{1.5pt}\arrayrulecolor{black}
Learning signal & scalar reward & reward plus the agent's actions, the oracle's output \\ \arrayrulecolor{black!12}\specialrule{0.3pt}{1.5pt}{1.5pt}\arrayrulecolor{black}
Credit assignment & group-relative advantage & diagnosis by the optimizer agent \\ \arrayrulecolor{black!12}\specialrule{0.3pt}{1.5pt}{1.5pt}\arrayrulecolor{black}
Update & gradient step & plan of edits \\ \arrayrulecolor{black!12}\specialrule{0.3pt}{1.5pt}{1.5pt}\arrayrulecolor{black}
Step size & learning rate, clipping, KL penalty & at most $b$ edits per plan \\ \arrayrulecolor{black!12}\specialrule{0.3pt}{1.5pt}{1.5pt}\arrayrulecolor{black}
Exploration & stochastic sampling of the model's outputs & $K$ alternative plans per round \\ \arrayrulecolor{black!12}\specialrule{0.3pt}{1.5pt}{1.5pt}\arrayrulecolor{black}
Acceptance & every step is taken & statistical gate on held-out tasks \\ \arrayrulecolor{black!12}\specialrule{0.3pt}{1.5pt}{1.5pt}\arrayrulecolor{black}
Optimizer state & moment estimates & ledger of past plans and outcomes \\
\bottomrule
\end{tabular}
\end{table}

\ding{182} \emph{Rollouts.} The target model runs the current harness $\Theta_t$ on a batch of $B$ training vulnerabilities that earlier rounds have not used, $n$ times per vulnerability.

\ding{183} \emph{Grading.} An execution oracle builds each patch in a fresh container and replays the PoV input under the sanitizer~\cite{asan}. To avoid rewarding patches that only mask the PoV, the training oracle also requires that variants of the PoV that crash the vulnerable program do not crash the patched one, that a LeakSanitizer run stays clean, and that the project's seed corpus still runs without error.

\ding{184} \emph{Diagnosis.} The optimizer is itself an agent. Once the whole batch is graded, it reads, for every attempt, the trace, the oracle's output and the developer's fix, and it can run commands in fresh training containers to test its hypotheses. It contrasts the attempts on the same vulnerability, which separates failures of the harness from failures that are specific to one run, and it targets problems that recur across vulnerabilities. The target model never sees the oracle's output or the fix. The optimizer runs in a sandbox without network access.

\ding{185} \emph{Planning.} The optimizer writes $K$ alternative plans $\Delta_t^{1},\dots,\Delta_t^{K}$, each of at most $b$ edits and each based on a different hypothesis about what limits the harness. A ledger records every round with its plans, their rationale, their gate statistics and whether they were accepted. The optimizer plans with the recent ledger entries, the full list of rejected plans and the files of its earlier rounds in context, so that it builds on successful edits and does not repeat failed ones.

\ding{186} \emph{Checks.} Each plan is applied to a copy $\Theta'_k=\Theta_t\oplus\Delta_t^{k}$. A filter removes edits that name files, functions or macros of the developer fixes the optimizer has read, which limits overfitting of the harness to the training vulnerabilities. Each new tool is then tested by running it inside multiple training containers. A plan that fails these checks returns to the optimizer for a bounded number of repairs.

\ding{187} \emph{Gate.} The current harness and every candidate run $m$ times on a fixed set $V$ of gate tasks, in the same pool and under the same conditions. Let $\hat{s}_\Theta(x)$ be the share of runs of $\Theta$ that resolve task $x$. The gain of candidate $k$ is
\begin{equation}
\hat{d}_k=\frac{1}{|V|}\sum_{x\in V}\big(\hat{s}_{\Theta'_k}(x)-\hat{s}_{\Theta_t}(x)\big),
\label{eq:gain}
\end{equation}
and $p_k$ is the $p$-value of an exact one-sided permutation test stratified by task. Among the candidates whose new tools fail on at most half of their calls, the gate selects $k^{\star}=\arg\max_k \hat{d}_k$ and updates the harness as
\begin{equation}
\Theta_{t+1}=\begin{cases}\Theta'_{k^\star} & \text{if } \hat{d}_{k^\star}>0,\ p_{k^\star}\le\alpha_s,\ \tilde{d}_{k^\star}>0,\ \tilde{p}_{k^\star}\le\alpha_c,\\ \Theta_t & \text{otherwise,}\end{cases}
\label{eq:gate}
\end{equation}
where $\tilde{d}$ and $\tilde{p}$ come from a confirmation on fresh runs of both harnesses. Selecting the best of $K$ candidates on the same runs inflates its gain, and the confirmation removes this bias. Outside Eq.~\eqref{eq:gate}, a last check rejects a candidate that makes the agent take markedly more steps or reach the step limit more often, unless a sign test over tasks supports the gain.

\textbf{No leakage.} Training projects are excluded from every evaluation benchmark. The optimizer never sees the traces of gate or evaluation runs. It receives only the gate statistics of its plans and the error messages of its new tools. Agents cannot download the upstream fix.

\section{Implementation and Experimental Setup}
\label{sec:setup}

\textbf{Implementation.} Harnesses are stored as YAML files. The runtime executes a harness with native tool calls inside an Apptainer container~\cite{apptainer} built from the task's image. It blocks downloads from the agents' commands, so that agents cannot fetch the upstream fix. An episode ends when the agent finishes, after 150 steps across all agents, or after 5{,}400\,s of execution time that excludes the time spent waiting for the model. For the methods that run in our runtime, episodes that end on this clock, and episodes that fail for infrastructure reasons, are run again and never count as failures. The optimizer is Claude Code\footnote{\url{https://code.claude.com/docs/en/overview}} with Claude Opus 5.5\footnote{\url{https://www.anthropic.com/claude/opus}}, run in headless mode in a sandbox without network access, with one call of at most one hour per round.

\begin{table}[!htbp]
    \caption{Datasets of vulnerabilities with patches. Training and gate tasks come from projects that no evaluation benchmark contains, except five gate projects that also appear in SEC-bench.}
    \label{tab:data}
    \centering\footnotesize
    \setlength{\tabcolsep}{3pt}
    \begin{tabular}{@{}llrr@{}}
    \toprule
    Data set & Role & Tasks & Projects \\
    \midrule
    ARVO~\cite{arvo} & training & 153 & 59 \\
    SEC-bench \texttt{oss} split & gate & 40 & 7 \\
    \midrule
    SEC-bench \texttt{cve} split~\cite{secbench} & evaluation & 200 & 29 \\
    CyberGym-E2E~\cite{cybergyme2e} & evaluation & 41 & 30 \\
    PatchBench~\cite{patchbench} & evaluation & 53 & 23 \\
    \bottomrule
    \end{tabular}
\end{table}

\textbf{Target models.} We build harnesses for Gemma-4-31B\footnote{\url{https://huggingface.co/google/gemma-4-31B-it}}, the smaller Gemma-4-E4B\footnote{\url{https://huggingface.co/google/gemma-4-E4B-it}} and GPT-OSS-120B\footnote{\url{https://huggingface.co/openai/gpt-oss-120b}}, a model of another family. Models are served with vLLM~\cite{vllm} 0.22.1 with a context of 131{,}072 tokens and the sampling parameters recommended by the model card (temperature 1.0, top-$p$ 0.95, top-$k$ 64). RQ5 and RQ6 add two models. Muse-Glimmer-30B\footnote{\url{https://huggingface.co/meta-models/Muse-Glimmer-30B}} serves both as an alternative optimizer and as a target model, and we also experimented with Nemotron-3.5\footnote{\url{https://huggingface.co/nvidia/NVIDIA-Nemotron-3.5-Lightning-30B-A3B-BF16}}.

\textbf{Build.} Every build starts from the base harness $\Theta_0$, a single agent with access to a shell and the repair instructions following the methodology from SEC-bench~\cite{secbench}. Training vulnerabilities come from ARVO~\cite{arvo}, a dataset of OSS-Fuzz vulnerabilities~\cite{ossfuzz}. We restrict them to projects that do not appear in any evaluation benchmark and keep at most eight per project, and use 153 vulnerabilities across 59 projects. Each round draws $B=24$ training vulnerabilities that earlier rounds have not used, starting a new pass over the pool once all have been used, and runs each $n=2$ times. The optimizer writes $K=2$ plans of at most $b=6$ edits. The gate tasks come from the \texttt{oss} split of SEC-bench, which is disjoint from the evaluation split. New tools are checked on three training images, and a plan that fails the checks gets up to two repairs. The gate uses $\alpha_s=0.3$ for the screen and $\alpha_c=0.1$ for the confirmation. The lenient screen keeps promising candidates, and the stricter confirmation on fresh runs decides. The step check rejects candidates that increase the mean number of steps by more than 30\%. Each build runs at most 12 rounds, and we evaluate the last accepted harness.

\textbf{Benchmarks.} We evaluate (Table~\ref{tab:data}) on the 200 CVE instances of SEC-bench~\cite{secbench}, on the 41 tasks of the development split of CyberGym-E2E~\cite{cybergym,cybergyme2e} whose developer fix passes the oracle in our containers, and on the 53 tasks of the development split of PatchBench~\cite{patchbench}. CyberGym-E2E and PatchBench are harder benchmarks, on which models resolve fewer tasks.

\textbf{Weight updates.} For RLVR, we train low-rank adapters~\cite{lora} (rank 32, $\alpha=64$) with a GRPO objective~\cite{grpo}, a learning rate of $10^{-5}$, two epochs per round and no KL penalty. The agent runs the base harness during training. For Gemma-4-31B, we report its evaluated checkpoint closest to the training budget of \rb{}, after 384 training episodes. For Gemma-4-E4B, we report its final checkpoint. For WSP (Section~\ref{sec:results}), the edits are designed from training episodes only, and they are frozen before any evaluation.

\textbf{Protocol and statistics.} Each method gets a single attempt per task, except ContraFix, which selects among its own candidates. We run \rb{}, the base harness and WSP in at least two independent passes, and we report the mean success rate over passes, the share of tasks for which the oracle accepts the patch. Differences between methods are paired by task, with 95\% confidence intervals from 10{,}000 bootstrap resamples of the tasks~\cite{bootstrap}. We count as training episodes the episodes whose traces or rewards a method learns from.

\providecommand{\pend}[1]{\textcolor{red}{#1}}

\section{Results}
\label{sec:results}

This section answers the six research questions stated in Section~\ref{sec:intro}.

\begin{table}[!htbp]
\caption{Success rate (\%) on the held-out benchmarks, averaged over two passes. All methods except the base harness learn from the same ARVO training pool. Bold: best method. $\Delta$: gain of \rb{} over the base harness (pp, paired by task); $^{*}$: 95\% confidence interval excludes zero.}
\label{tab:main}
\centering\footnotesize
\setlength{\tabcolsep}{2.6pt}
\begin{tabular}{@{}llccc@{}}
\toprule
& Method & SEC-bench & CyberGym-E2E & PatchBench \\
\midrule
\multirow{7}{*}{\rotatebox{90}{Gemma-4-31B}}
& Base harness & 45.5 & 40.2 & \phantom{0}5.3 \\
& EvoRepair & 50.5 & 53.7 & 15.1 \\
& ContraFix & 64.3 & 41.5 & \textbf{20.8} \\
& RLVR & 57.5 & 53.7 & 17.0 \\
& WSP & 59.8 & 48.8 & 17.9 \\
\arrayrulecolor{black!15}\cmidrule(l){2-5}\arrayrulecolor{black}
& \rb{} & \textbf{77.7} & \textbf{70.7} & 18.5 \\
& \quad $\Delta$ vs.\ base & +32.2$^{*}$ & +30.5$^{*}$ & +13.0$^{*}$ \\
\midrule
\multirow{7}{*}{\rotatebox{90}{Gemma-4-E4B}}
& Base harness & \phantom{0}4.8 & \phantom{0}4.9 & \phantom{0}0.0 \\
& EvoRepair & \phantom{0}6.0 & 12.2 & \phantom{0}3.8 \\
& ContraFix & 33.5 & \textbf{24.4} & \phantom{0}5.7 \\
& RLVR & \phantom{0}4.5 & \phantom{0}2.4 & \phantom{0}1.9 \\
& WSP & 11.8 & 12.2 & \phantom{0}4.7 \\
\arrayrulecolor{black!15}\cmidrule(l){2-5}\arrayrulecolor{black}
& \rb{} & \textbf{40.8} & 17.1 & \textbf{10.4} \\
& \quad $\Delta$ vs.\ base & +36.1$^{*}$ & +12.2$^{*}$ & +10.6$^{*}$ \\
\midrule
\multirow{3}{*}{\rotatebox{90}{\scriptsize GPT-OSS}}
& Base harness & 59.5 & 58.7 & 16.0 \\
& \rb{} & \textbf{91.8} & \textbf{73.9} & \textbf{28.3} \\
& \quad $\Delta$ vs.\ base & +32.2$^{*}$ & +15.2$^{*}$ & +12.3$^{*}$ \\
\bottomrule
\end{tabular}
\end{table}

\subsection{RQ1: Comparison with Harness Builders and Weight Updates}

Table~\ref{tab:main} compares \rb{} with the base harness and with four methods that learn from the same training pool. EvoRepair and ContraFix are the state of the art in harness building for vulnerability repair. RLVR and WSP change the model weights instead of the harness.

\begin{table*}[t]
   \caption{SEC-bench success rate (\%) by vulnerability class, as reported by the sanitizer, and by the size of the developer's fix in changed lines. Bold: best method.}
   \label{tab:classes}
   \centering\footnotesize
   \setlength{\tabcolsep}{4.2pt}
   \begin{tabular}{@{}lrcccccc@{\hspace{10pt}}!{\color{black!20}\vrule width 0.4pt}@{\hspace{10pt}}cccccc@{}}
   \toprule
   & & \multicolumn{6}{c}{Gemma-4-31B} & \multicolumn{6}{c}{Gemma-4-E4B} \\
   \cmidrule(l{4pt}r{14pt}){3-8}\cmidrule(l{-4pt}){9-14}
   & $n$ & Base & EvoRepair & ContraFix & RLVR & WSP & \rb{} & Base & EvoRepair & ContraFix & RLVR & WSP & \rb{} \\
   \midrule
   Heap overflow & 74 & 43.9 & 48.6 & 66.2 & 54.1 & 61.5 & \textbf{77.9} & \phantom{0}6.8 & 12.2 & 40.5 & \phantom{0}4.1 & 11.5 & \textbf{47.3} \\
   Null or wild pointer & 71 & 41.5 & 54.9 & 63.4 & 54.9 & 59.9 & \textbf{77.0} & \phantom{0}1.4 & \phantom{0}1.4 & 31.0 & \phantom{0}1.4 & \phantom{0}9.2 & \textbf{40.8} \\
   Use after free & 17 & 44.1 & 47.1 & 47.1 & 58.8 & 35.3 & \textbf{70.6} & \phantom{0}5.9 & \phantom{0}5.9 & 17.6 & \phantom{0}5.9 & 11.8 & \textbf{27.4} \\
   Memory leak & 13 & 61.5 & 30.8 & 69.2 & 69.2 & 73.1 & \textbf{79.5} & \textbf{15.4} & \phantom{0}7.7 & \textbf{15.4} & \phantom{0}0.0 & \phantom{0}3.8 & \phantom{0}5.1 \\
   Stack overflow & 10 & 55.0 & 40.0 & 60.0 & \textbf{90.0} & 70.0 & 86.7 & \phantom{0}0.0 & \phantom{0}0.0 & \textbf{50.0} & 20.0 & 25.0 & \textbf{50.0} \\
   Other & 15 & 53.3 & 66.7 & 73.3 & 53.3 & 60.0 & \textbf{80.0} & \phantom{0}3.3 & \phantom{0}0.0 & 33.3 & 13.3 & 23.3 & \textbf{48.9} \\
   \midrule
   Fix of 1--6 lines & 87 & 56.3 & 51.7 & 69.0 & 65.5 & 69.5 & \textbf{82.4} & \phantom{0}4.0 & \phantom{0}9.2 & 34.5 & \phantom{0}4.6 & 14.4 & \textbf{43.7} \\
   Fix of 7--20 lines & 79 & 39.9 & 54.4 & 58.2 & 53.2 & 53.8 & \textbf{72.2} & \phantom{0}6.3 & \phantom{0}5.1 & 36.7 & \phantom{0}5.1 & \phantom{0}8.9 & \textbf{37.1} \\
   Fix of $>$20 lines & 34 & 30.9 & 38.2 & 64.7 & 47.1 & 48.5 & \textbf{78.4} & \phantom{0}2.9 & \phantom{0}0.0 & 23.5 & \phantom{0}2.9 & 11.8 & \textbf{42.2} \\
   \bottomrule
   \end{tabular}
\end{table*}

\textbf{SEC-bench.} \rb{} gives the highest success rate for both Gemma-4 models. It raises Gemma-4-31B from 45.5\% to 77.7\%, a gain of 32.2 percentage points (pp) and a relative improvement of 71\%, and it outperforms ContraFix, the strongest other method, by 13.2\,pp (21\% relative). Since all methods learn from the same training pool, this gap comes from what each method is able to change. ContraFix and EvoRepair add past experiences or skills to the context of the agent, whereas \rb{} also changed the tools of the agent and the control of its episodes. RLVR and WSP, which update the model weights, remain about 20\,pp below \rb{}. With the same training data, a harness adapted to the model can therefore take it further than updating its weights, since the harness can compensate for the weaknesses of the model, for example with a tool that applies edits reliably or a guard that interrupts its loops. \\
The gap is wider for the smaller Gemma-4-E4B. Its base harness resolves only 4.8\% of SEC-bench, and \rb{} raises this rate to 40.8\%, more than eight times as many vulnerabilities. The low success rate of a small model with a base harness therefore does not necessarily reflect its ability to repair vulnerabilities, and a harness adapted to the model can unlock much of this ability. ContraFix comes closest, 7.3\,pp below \rb{}, and the other methods stay below 12\%. RLVR was not able to obtain significant gains on Gemma-4-E4B. With a base-harness success rate below 5\%, almost every training episode earns no reward, so the policy gradient has little to learn from. \rb{}, in contrast, can still learn from these failed episodes, because its optimizer reads their traces. \\
On GPT-OSS-120B, from another model family, \rb{} raised the success rate from 59.5\% to 91.8\% (54\% relative). For this model, the optimizer found that the agent repeatedly corrupted the source files with multi-line \texttt{sed} edits. It added an \texttt{apply\_patch} tool to the harness, which gave the model the interface it expected.

\begin{figure}[t]
   \centering
   \includegraphics[width=\columnwidth]{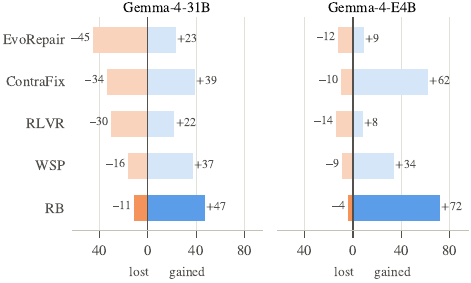}
   \caption{SEC-bench tasks that each method solves and the base harness does not (gained), and tasks that the base harness solves and the method does not (lost).}
   \label{fig:flips}
\end{figure}

\textbf{Harder benchmarks.} The gains carry over to CyberGym-E2E and PatchBench. \rb{} improves Gemma-4-31B significantly on both, by 30.5\,pp (76\% relative) and 13.0\,pp (3.5 times its base-harness rate), Gemma-4-E4B by 12.2\,pp and 10.6\,pp, where its base harness resolves no PatchBench task, and GPT-OSS-120B by 15.2\,pp and 12.3\,pp. This indicates that the harnesses built by \rb{} capture repair procedures that are not specific to one benchmark and are able to provide gains on more difficult tasks. On these smaller benchmarks, the methods are closer. For Gemma-4-31B, \rb{} has the best rate on CyberGym-E2E, and for Gemma-4-E4B, it has the best rate on PatchBench. No other method was significantly better than \rb{} on either benchmark.

\textbf{Where the gains come from.} Compared with the base harness, \rb{} solves many new vulnerabilities and rarely fails on those that the base harness already solved (Fig.~\ref{fig:flips}). On Gemma-4-31B, it solves 47 vulnerabilities that the base harness does not, and fails on 11 that the base harness solves. ContraFix gains 39 tasks but also loses 34, so much of what it adds is offset by tasks that the base harness already solved. On Gemma-4-E4B, \rb{} gains 72 tasks and loses 4. This indicates that, unlike ContraFix, \rb{} builds a stronger harness with little degradation of what the model could already do. This likely stems from the gate, which keeps a change only if it improves the success rate on held-out tasks, limiting overfitting. The gains also span different vulnerability classes and the difficulty of the fix (Table~\ref{tab:classes}). \rb{} has the highest or a tied success rate in five of the six classes for both Gemma-4 models, so the gains do not appear to be specific to one type of vulnerability. Its gain is largest where the developer's fix is longest. For fixes of more than 20 lines, it raises Gemma-4-31B from 30.9\% to 78.4\%, so the built harness also helps on the hardest vulnerabilities.

\textbf{Weight-space programming.} To test how far a targeted change of the weights can go, an agent edited the weights of each model directly, following WSP~\cite{wsp}. The agent read the training traces, located the decision at which unsolved episodes go wrong and wrote a closed-form edit for it, without gradient training. For Gemma-4-31B, 73 of 88 unsolved training episodes ended before editing a file and cycled through \texttt{sed -n}, \texttt{cat} and \texttt{grep} commands instead. The model weight editor agent fitted a detector for this state on the input of the last MLP layer and repurposed four of its hidden units, 64{,}512 weights in total, or 0.0002\% of the 32.7 billion weights of the model. When the detector fires, these units push the next command from viewing code with \texttt{sed -n} to editing it with \texttt{sed -i}. For Gemma-4-E4B, the agent found the direction in the activations that makes the model end the episode as soon as it has reproduced the crash, and removed it from the weights. It also repurposed 14 hidden units so that, when a build fails, the model discards its broken edits with \texttt{git checkout .} and starts again from the original code. Both models run the base harness with their edited weights, so any change in their success rate comes from the edit alone.

WSP raises Gemma-4-31B by 14.2\,pp on SEC-bench, since breaking the reading loop lets the model apply fixes that it had already identified. It raises Gemma-4-E4B by only 7.0\,pp, because most failed attempts of this model leave the vulnerability in place, and a small weight edit is unlikely to teach it a correct fix, as also suggested by the results from RLVR. The harness built by \rb{} instead makes the model test and revise its patch until the PoV no longer triggers the vulnerability (Section~\ref{sec:rq3}).

\begin{rqanswer}\emph{Answer to RQ1:} Overall, \rb{} built the strongest harness for the tested models, with gains of 26\% to 750\% relative to the base harness (10.6 to 36.1\,pp), and it outperformed ContraFix and EvoRepair by 24\% and 66\% relative on average on the two Gemma-4 models. Learning the whole harness, including its tools and the control of its episodes, went further than adding past experiences to the context of the agent, and further than updating the model weights on the same data, since the harness can compensate for the specific weaknesses of each model. The gains generalized to the harder benchmarks, where no other method was significantly better, and they came with few regressions on the vulnerabilities that the model already solved with the base harness.\end{rqanswer}

\subsection{RQ2: Learning Efficiency and Build Dynamics}

\begin{figure}[t]
\centering
\includegraphics[width=0.95\columnwidth]{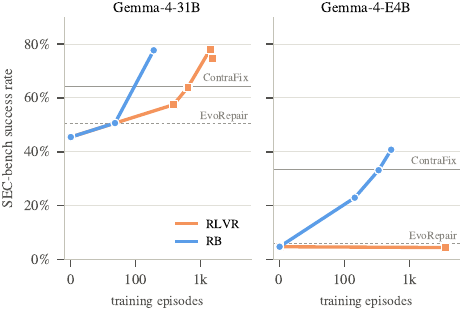}
\caption{SEC-bench success rate after each accepted round of \rb{} and after each evaluated round of RLVR, against the number of training episodes (log scale). Grey lines mark the two harness builders, ContraFix (solid) and EvoRepair (dashed).}
\label{fig:curves}
\end{figure}

Figure~\ref{fig:curves} compares the success rate of \rb{} and RLVR on the held-out SEC-bench evaluation set as a function of the number of training episodes. On Gemma-4-31B, \rb{} reaches 77.7\% after 190 training episodes, whereas RLVR needs 1{,}408 episodes, 7.4 times as many, to reach the same rate. We attribute this efficiency to the learning signal. The optimizer of \rb{} reads the whole trace and can identify a failure from a few episodes, whereas RLVR, trained on the binary reward of the benchmark, only learns whether each episode succeeded. On Gemma-4-E4B, \rb{} reaches 40.8\% after 525 training episodes, while RLVR was not able to obtain significant gains even after 3{,}584 episodes. When the target model rarely succeeds, the traces can therefore provide a learning signal that the binary reward does not, without a reward designed in advance for the failures of the model. Over the whole build for Gemma-4-31B, the optimizer worked for 85 minutes, whereas RLVR trains the model weights on GPUs for about three hours in every round.

\begin{figure}[t]
\centering
\includegraphics[width=0.8\columnwidth]{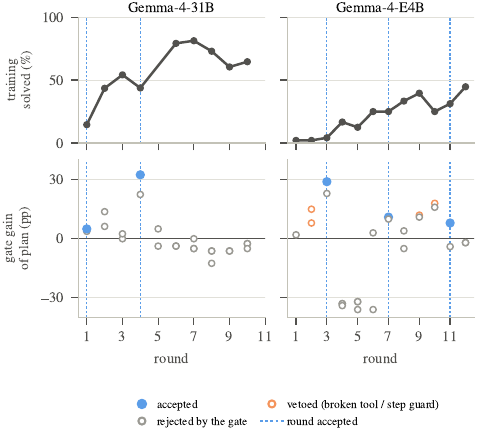}
\caption{Progress of the \rb{} builds for Gemma-4-31B and Gemma-4-E4B over rounds. Top: success rate of the current harness on the training episodes of each round. Bottom: gain $\hat d$ of each candidate plan over the current harness on the gate tasks, in percentage points (Eq.~\ref{eq:gain}).}
\label{fig:build}
\end{figure}

\textbf{Build dynamics.} Figure~\ref{fig:build} shows how the two Gemma-4 builds progress. On the training vulnerabilities, the success rate of the current harness rises from 15\% to about 80\% for Gemma-4-31B and from 2\% to 45\% for Gemma-4-E4B. A plan that helps on the training vulnerabilities may overfit them and not help on other tasks, so the gate checks each plan on held-out tasks. It rejected 17 of the 22 rounds. Some of the rejected plans had a small gain that was not significant, and others lowered the success rate. In rounds 4 to 6 of the build for Gemma-4-E4B, for example, five plans lowered it by more than 30\,pp on the gate tasks. After round 4 of the build for Gemma-4-31B, every further plan failed the gate, and the harness stayed unchanged even though the optimizer kept proposing edits. The checks also matter. In round 2 of the build for Gemma-4-E4B, a plan with a gate gain of 15\,pp added a \texttt{fixcheck} tool that failed in 590 of its 611 calls. The checks rejected it, and the optimizer repaired the tool in the next round, where it produced the largest gain of the build. Without the checks, the build would have kept a tool that almost never worked.

\begin{figure*}[t]
\centering
\includegraphics[width=\textwidth]{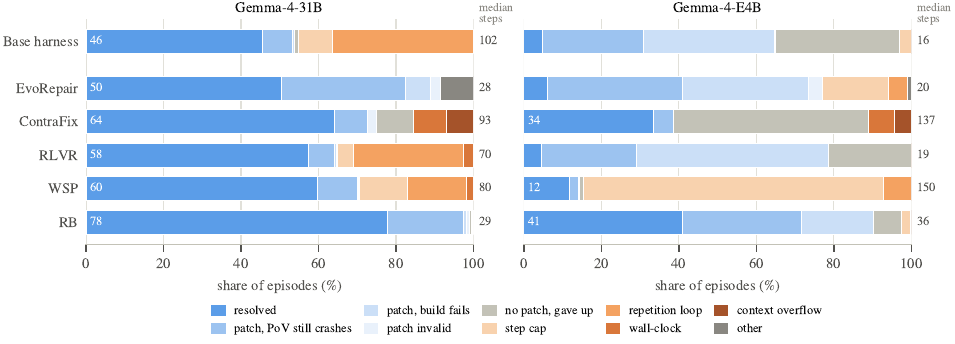}
\caption{How the SEC-bench episodes of each method end. Blue outcomes submit a patch, grey outcomes stop without one, and orange outcomes are unfinished episodes, which a step, time or context limit ended before the agent submitted a patch. The right column gives the median number of steps.}
\label{fig:behavior}
\end{figure*}

\begin{figure}[t]
\centering
\includegraphics[width=\columnwidth]{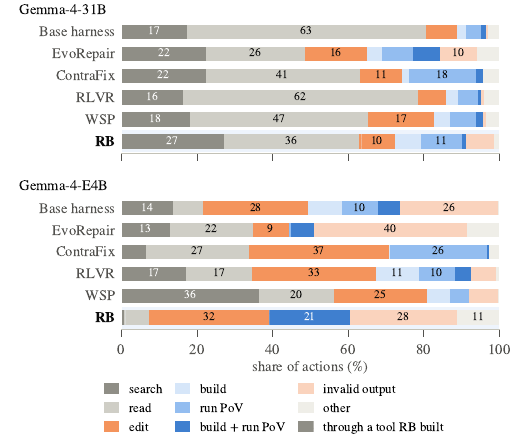}
\caption{What the agent spent its actions on in its SEC-bench episodes, all passes pooled. Hatching marks actions made through a tool that \rb{} built (Appendices~\ref{app:h31b} and~\ref{app:he4b}). Invalid outputs are malformed tool calls, which the ContraFix traces do not record.}
\label{fig:actions}
\end{figure}

\begin{rqanswer}\emph{Answer to RQ2:} \rb{} was able to reach the same success rate as RLVR on Gemma-4-31B from 7.4 times fewer training episodes, without training the model weights. It also improved Gemma-4-E4B, for which RLVR was not able to obtain significant gains. This indicates that updating the harness provides a richer learning signal than updating the model weights alone, in particular when the target model rarely succeeds.\end{rqanswer}

\subsection{RQ3: Changes in Agent Behavior}
\label{sec:rq3}

Figure~\ref{fig:behavior} shows how the SEC-bench episodes ended with each of the learned harnesses or models. With the base harness, the two Gemma-4 models failed in different ways. Gemma-4-31B often got stuck. In 45\% of its episodes, a step or time limit ended the run before the agent submitted a patch, and we call these episodes unfinished. Most of them (37\% of all episodes) were repetition loops, in which the agent repeated a command 23 times per episode on average. Gemma-4-E4B instead gave up early, after a median of 16 steps, either with a wrong patch (60\%) or with no patch at all (32\%). A harness that suits one of these models is therefore unlikely to suit the other.

\textbf{Gemma-4-31B.} \rb{} almost eliminated the unfinished episodes, from 45\% to 0.5\%. The agent repeated a command 1.7 times per episode instead of 23, and it finished after a median of 29 steps instead of 102. Since these episodes were the main failure of this model, removing them accounts for a large part of its gain. The model was not able to escape its loops by itself, whereas the built harness detects them from outside, interrupts them and lets the agent restart with a note on what it learned. The weight updates addressed the same failure less effectively. RLVR and WSP reduced the unfinished episodes only to 35\% and 30\%, and repetition loops remained their main failure, even though WSP doubled the share of commands that edit code, as it was designed to do. The harness builders left other failures. EvoRepair always submitted a patch, but 41\% of its patches failed the oracle, and the agents of ContraFix ran out of time or exceeded the context window in 15\% of the episodes.

\textbf{Gemma-4-E4B.} On this model, the weight updates changed how the episodes ended more than whether the patches were correct. RLVR lowered the share of episodes without a patch from 32\% to 22\%, but most of the additional patches did not build. WSP kept the model working, but 78\% of its episodes then reached the step limit. ContraFix withheld the patches it could not verify, so half of its episodes ended without a patch. With \rb{}, the share of episodes without a patch fell from 32\% to 7\%, the agent rebuilt the project and reran the PoV 9.4 times per episode instead of 5.1, and 41\% of the episodes ended with a correct patch. The main failure of this model was to stop before verifying its patch, and the built harness addressed it by making verification a single tool call and a condition for finishing. The harness thus improved the success rate by making the model test and revise its patches, without giving it new knowledge about the vulnerabilities. Across methods, the median patch stayed small, at four to five changed lines on Gemma-4-31B and two to eight on Gemma-4-E4B, so the gains did not come from larger patches.

\textbf{Actions.} Fig.~\ref{fig:actions} shows what the agent spent its actions on. With the base harness, Gemma-4-31B spent 63\% of its actions reading files and 8\% building the project or running the PoV. The harness that \rb{} built replaced much of this reading with targeted search and verification: reading fell to 36\% of the actions, search rose from 17\% to 27\%, and building and running the PoV rose to 19\%. RLVR left the action mix almost unchanged. For Gemma-4-E4B, the tools that \rb{} built became the agent's main interface, with 54\% of its actions going through them. Its \texttt{fixcheck} tool, which rebuilds the project and reruns the PoV in one call, accounted for 21\% of the actions and replaced the separate build and PoV commands of the base harness. For both Gemma-4 models, the built harnesses moved the actions of the agent from reading code to checking its patch on the program.

\begin{table*}[t]
\caption{The accepted rounds of the two Gemma-4 builds: what the optimizer diagnosed in the training episodes, the edits it made and the gain $\tilde d$ of the harness on fresh gate runs. Diagnoses are condensed from the ledger.}
\label{tab:updates}
\centering\footnotesize
\setlength{\tabcolsep}{4pt}
\begin{tabular}{@{}l c L{0.47\textwidth} L{0.26\textwidth} r@{}}
\toprule
Model & Round & Diagnosis of the optimizer & Accepted edits & $\tilde d$ (pp) \\
\midrule
Gemma-4-31B & 1 & About 20 of 47 episodes re-read code in loops and never edit. Eleven repeat one command 68 to 130 times in a row. Edits fail on tabs and spaces in the source. & rewrite instructions, add \texttt{replace} tool, loop guard 3, two restarts & +12.5 \\
Gemma-4-31B & 4 & The model writes comments inside its commands. Episodes with two or more comment blocks among their first 20 commands were fixed in 70\% of the cases and never looped, against 43\% fixed and 28\% loops without. & rewrite instructions: a hypothesis comment before every command, root cause and fix within about 15 commands & +32.5 \\
\midrule
Gemma-4-E4B & 3 & Paths from the stack trace are used relative to the wrong directory, multi-line \texttt{sed} edits silently match nothing, and the agent reruns the PoV without rebuilding. & rewrite instructions, add \texttt{fixcheck}, \texttt{showsrc} and \texttt{editsrc}, add a memory note & +28.0 \\
Gemma-4-E4B & 7 & 26 of the 28 failed builds end with \texttt{finish} right after ``build failed''. The line editor duplicates the lines around the replaced range. & repair \texttt{editsrc} and \texttt{fixcheck} & +13.0 \\
Gemma-4-E4B & 11 & Most failed builds come from structural damage by the agent's own edits, which it repeats because it never sees what its edits changed. & revise \texttt{fixcheck}, add a memory note on recovering from failed builds & +12.0 \\
\bottomrule
\end{tabular}
\end{table*}

The ledger shows how the optimizer arrived at these harnesses (Table~\ref{tab:updates}). Each accepted edit addressed a failure that the optimizer had counted in the training episodes. The clearest case is round 4 of the build for Gemma-4-31B. The optimizer noticed that the model reasoned only in the comments of its commands, measured that episodes with such comments succeeded more often and never looped, and made these comments mandatory. This single edit raised the gate success rate by 32.5\,pp. The binary reward does not carry this information, because it does not include what the agent wrote.

\textbf{The built harnesses.} The two harnesses that \rb{} built explain these changes, and they differ between the models (Fig.~\ref{fig:harness}). For Gemma-4-31B, \rb{} kept a single agent and changed how it works. The agent must write its current hypothesis as a comment before each command, read each region of code only once, and state the root cause and the fix within about 15 commands. A \texttt{replace} tool edits an exact block of lines and refuses ambiguous matches. A loop guard interrupts the agent after three repeated commands, and the agent can restart with a fresh context and a note to itself, at most twice. For Gemma-4-E4B, \rb{} gave the agent three tools. \texttt{showsrc} prints numbered source lines, \texttt{editsrc} replaces a numbered range of lines, and \texttt{fixcheck} rebuilds the project, runs the PoV and prints one of three verdicts. The instructions forbid finishing until \texttt{fixcheck} reports a local fix, and two memory notes explain how to read each type of sanitizer report and how to recover from a failed build. Each harness addresses the failure mode of its own model. The first stops the loops of Gemma-4-31B, and the second keeps Gemma-4-E4B working until its patch builds and the PoV no longer triggers the vulnerability.

\begin{figure}[!htbp]
\centering
\begin{minipage}{\columnwidth}
\scriptsize
\textbf{Gemma-4-31B} (one agent, tool \texttt{replace}, loop guard 3, 2 restarts)\vspace{1pt}
\begin{promptbox}[left=1pt,right=0pt]
\begin{verbatim}
Think in comments: start EVERY command block with one or
two `# ' lines: what the last output told you and what
you check next and why. [...]
1. Run `rb-repro' once to see the crash stack. [...]
   Read each region ONCE: never re-run a command you
   already ran, its output is still above.
2. Within about 15 commands write `# ROOT CAUSE: ...'
   and `# FIX: ...' comments and act on them. [...]
Getting unstuck: if you get a repeated-command warning
[...] call the `restart' tool with a note: crash
location, your root-cause hypothesis, what you
changed and why it failed, what to try next.
\end{verbatim}
\end{promptbox}
\vspace{2pt}
\textbf{Gemma-4-E4B} (one agent, tools \texttt{showsrc}, \texttt{editsrc}, \texttt{fixcheck}, 2 memory notes)\vspace{1pt}
\begin{promptbox}[left=1pt,right=0pt]
\begin{verbatim}
- `fixcheck' rebuilds the project and runs the PoC. It
  prints "result: build failed" (with the compiler
  errors), "result: still crashes" (with the sanitizer
  stack) or "result: fixed locally".
4. Never finish unless the latest `fixcheck' printed
   "result: fixed locally". [...] If one approach fails
   3 times, undo it with `git checkout -- FILE' and try
   a different place or approach.
memory: When fixcheck prints 'result: build failed',
  assume your edit caused it and never finish in that
  state. [...]
\end{verbatim}
\end{promptbox}
\end{minipage}
\caption{Excerpts of the two harnesses that \rb{} built, quoted from their instructions.}
\label{fig:harness}
\end{figure}

\textbf{Two examples.} In both examples below, the base agent had the right first idea, and the built harness made the model act on this idea and verify it.

The first example is a 2022 vulnerability in mruby, where a NULL class pointer reaches \texttt{mrb\_obj\_is\_kind\_of}. With the base harness, Gemma-4-31B reproduced the crash, opened the crashing function and read its call site within a few steps. It then printed the same 30 lines of \texttt{src/object.c}, or the same macro definition, 133 times in a row until the step limit ended the run, and it never edited a file. With the built harness, the same model wrote its hypothesis before each command, traced the NULL pointer back to the \texttt{super} path of the interpreter and stated its root cause and fix in two comments. It added a one-line guard with \texttt{replace}, rebuilt the project and saw the crash turn into the interpreter's own type error after 17 steps. The built harness resolved this vulnerability in all three of its passes, and the base harness in none.

The second example, a 2021 vulnerability in GPAC, is a NULL-pointer dereference in a function that exports media tracks. With either harness, Gemma-4-E4B first guarded \texttt{esd->decoderConfig} in the condition at the crashing line. This guard was incomplete, because the \texttt{else} branch of the same block dereferences the pointer again. The base agent reran the PoV four times without rebuilding the project, so each ``still crashes'' came from the old binary. Its \texttt{sed} rewrites then duplicated code and broke the build, and it finished after 23 steps, noting that ``repeated verification with \texttt{rb-repro} indicated the crash persists''. With the built harness, \texttt{fixcheck} rebuilt the project, reran the PoV after the first edit and reported that the crash remained. The model then viewed the whole block with \texttt{showsrc}, found the unguarded \texttt{else} branch, guarded it with \texttt{editsrc}, and stopped after 11 steps when \texttt{fixcheck} reported ``fixed locally''. The built harness resolved this vulnerability in all three of its passes, and the base harness in none.

\begin{rqanswer}\emph{Answer to RQ3:} \rb{} was able to adapt the harness to the main weakness of each model. It removed the repetition loops of Gemma-4-31B, which the model could not escape by itself and which the weight updates only reduced, and it made Gemma-4-E4B test and revise its patch until the PoV no longer triggered the vulnerability. The weight updates mostly changed how the episodes of Gemma-4-E4B ended, not whether its patches were correct, which suggests that the harness is a more direct lever on these failures.\end{rqanswer}

\subsection{RQ4: Contribution of the Harness Components}

\begin{figure}[!htbp]
\centering
\includegraphics[width=\columnwidth]{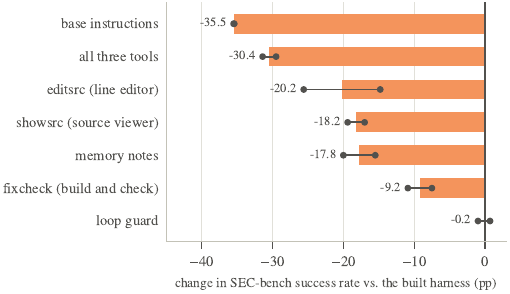}
\caption{Change in the SEC-bench success rate of Gemma-4-E4B when one component is removed from its built harness. Bars show the mean of two evaluation passes, and the dots show each pass.}
\label{fig:ablation}
\end{figure}

\textbf{Harness components.} We removed one component at a time from the harness built for Gemma-4-E4B (Fig.~\ref{fig:ablation}). When a tool is removed, the instructions that name it point to the plain shell command instead. Every component that \rb{} added contributes to the gain. Restoring the base instructions costs 35.5\,pp, almost the whole gain of 36.1\,pp, and removing all three tools costs 30.4\,pp. Removing \texttt{editsrc} costs 20.2\,pp, \texttt{showsrc} 18.2\,pp and \texttt{fixcheck} 9.2\,pp, and the two memory notes are worth 17.8\,pp. The loop guard has no measurable effect for this model. The losses of the single components add up to far more than the whole gain, so the components work together. The instruction to finish only after \texttt{fixcheck} reports a fix, for example, helps only if the agent can also view and edit the source reliably.

\textbf{Developer fix.} The optimizer reads the developer's fix of each training vulnerability during diagnosis. To test whether the gain depends on these fixes, we ran two four-round builds for Gemma-4-E4B in which the optimizer reads the traces and oracle outputs but no fix, with the same training pool, gate tasks and optimizer. These builds reached 20.2\% on SEC-bench, about five times the 4.0\% of the base harness in the same evaluation runs (+16.2\,pp), and close to the 23.0\% that the full build reached after its first accepted round. \rb{} therefore does not need the developer fixes to improve the harness.

\begin{rqanswer}\emph{Answer to RQ4:} The instructions, each of the three tools and the memory notes that \rb{} added all contributed to the gain, and they worked together. This indicates that the gain came from adapting several components of the harness jointly, which goes beyond adding knowledge to the context of the agent. \rb{} also did not need the developer fixes: without them, its builds still reached about five times the base-harness rate, which suggests that the optimizer learns mostly from the traces of the target model.\end{rqanswer}

\subsection{RQ5: Role of the Optimizer and of the Target Model}

\textbf{Optimizer.} \rb{} also worked with other optimizers than Claude Opus 5.5 (Table~\ref{tab:optimizer}). With Muse-Glimmer-30B as the optimizer, it raised Gemma-4-31B from 41.5\% to 68.2\% on SEC-bench, a relative improvement of 64\%, and the same harness also improved this model on CyberGym-E2E and PatchBench. When Gemma-4-31B optimized its own harness, the gain fell to 6.5\,pp, which suggests that the optimizer needs to be stronger than the target model to diagnose its failures.

\begin{table}[!htbp]
\caption{SEC-bench success rate (\%) of harnesses built by \rb{} with other optimizers and target models. Base-harness and \rb{} rates of each row come from the same evaluation runs.}
\label{tab:optimizer}
\centering\footnotesize
\setlength{\tabcolsep}{2.2pt}
\begin{tabular}{@{}llccr@{}}
\toprule
Target model & Optimizer & Base & \rb{} & $\Delta$ \\
\midrule
Gemma-4-31B & Muse-Glimmer-30B & 41.5 & \textbf{68.2} & +26.7 \\
Gemma-4-31B & Gemma-4-31B (itself) & 41.5 & 48.0 & +6.5 \\
GPT-OSS-120B & Muse-Glimmer-30B & 60.2 & 63.9 & +3.8 \\
Muse-Glimmer-30B & Muse-Glimmer-30B & 87.5 & 87.5 & 0.0 \\
\bottomrule
\end{tabular}
\end{table}

\textbf{Target model.} The choice of optimizer mattered more for GPT-OSS-120B. Muse-Glimmer-30B improved it by only 3.8\,pp, whereas Claude Opus 5.5 improved it by 32.2\,pp (Table~\ref{tab:main}). For Muse-Glimmer-30B as the target model, the base harness already resolved 87.5\% of SEC-bench, close to the ceiling of the benchmark, and \rb{} did not improve it further.

\begin{rqanswer}\emph{Answer to RQ5:} The gains of \rb{} were not tied to a specific optimizer model: Muse-Glimmer-30B also built a harness that substantially improved Gemma-4-31B. The gains were smaller when the model optimized its own harness, which indicates that \rb{} benefits from an optimizer that is stronger than the target model.\end{rqanswer}

\subsection{RQ6: Transfer Across Models}

\begin{figure}[!htbp]
\centering
\includegraphics[width=\columnwidth]{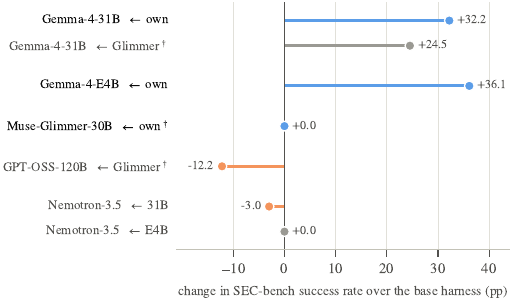}
\caption{Change in the SEC-bench success rate (percentage points) when a model runs a harness that \rb{} built, relative to its base harness. Each row names the model that runs the harness and the model the harness was built for; blue rows are a model's own harness. $^\dagger$Built with Muse-Glimmer-30B as the optimizer.}
\label{fig:transfer}
\end{figure}

A harness that works well for one model does not necessarily work for another (Fig.~\ref{fig:transfer}). The harness built for Muse-Glimmer-30B lowers the success rate of GPT-OSS-120B by 12.2\,pp. The harness built for Gemma-4-31B changes that of Nemotron-3.5 by $-3.0$\,pp, and the harness built for Gemma-4-E4B leaves it unchanged.

A transferred harness can still help a weaker model. The harness built for Muse-Glimmer-30B raises Gemma-4-31B by 24.5\,pp, because it carries general repair procedures that the weaker model lacks. The harness that \rb{} built for Gemma-4-31B itself, however, raises it by 32.2\,pp, more than any transferred harness. Since a build needs only a few hundred training episodes (RQ2), building the harness for each model is practical and gives larger gains than reusing one.

\begin{rqanswer}\emph{Answer to RQ6:} A harness adapted to one model did not necessarily help another, and it could even lower its success rate. Each model gained most from its own harness, which supports building a harness for each model, as \rb{} does from a few hundred training episodes.\end{rqanswer}

\providecommand{\pend}[1]{\textcolor{red}{#1}}

\section{Discussion}
\label{sec:discussion}

\textbf{Why a harness learns from fewer episodes.} The edits that \rb{} accepted encode procedures, such as verifying the patch before finishing or reading a region of code only once. With a binary reward, RLVR can learn such a procedure only if the model happens to follow it in enough rewarded episodes, and the reward does not say which part of an episode mattered. The optimizer of \rb{} reads the episode and names the step that went wrong, so one round can address a failure that the model rarely avoids on its own. This is consistent with both results of RQ2. On Gemma-4-31B, RLVR eventually reaches the level of \rb{}, but it needs 7.4 times as many episodes. On Gemma-4-E4B, almost no episode is rewarded, and RLVR was not able to obtain significant gains. We trained RLVR on the binary execution reward of the benchmarks. A shaped reward could narrow the gap on Gemma-4-E4B, but it would have to be designed for the failure modes of each model, which \rb{} identifies automatically from the traces.

\textbf{What the optimizer contributes.} A reader may ask whether \rb{} transfers the knowledge of a strong optimizer to the target model rather than adapting the harness. Three observations argue against this. First, the optimizer never sees an evaluation task, a filter removes edits that name the files or functions of the training fixes it has read, and builds whose optimizer reads no fix at all still reach about five times the base-harness rate (RQ4). Second, the built harnesses contain general procedures and tools rather than repair knowledge (Section~\ref{sec:rq3}), and removing the tools or the instructions removes most of the gain (RQ4). Third, \rb{} with Muse-Glimmer-30B as the optimizer raises Gemma-4-31B by 26.7\,pp (RQ5), a gain of the same order as with Claude Opus 5.5. The optimizer must still understand the failures it reads. When Gemma-4-31B optimizes its own harness, the gain falls to 6.5\,pp.

\textbf{Cost.} A build needs only inference with the target model and a few calls to the optimizer. The build for Gemma-4-31B made 13 optimizer calls that took 85 minutes and cost about \$22 at API prices, and the build for Gemma-4-E4B made 12 calls (98 minutes, about \$27). RLVR instead trains the model weights on GPUs in every round.

\textbf{Implications for defenders.} A defensive team that adopts a new model can build its harness with a few hundred training episodes on vulnerabilities from its own projects, without access to the weights and without GPU training. When its projects have few known vulnerabilities, training tasks could also be generated automatically, for example by injecting realistic vulnerabilities into their code~\cite{lbath2026aviator,lbath2026toward}. The built harness is a short, readable file. The team can inspect every tool and instruction before deployment, which is harder for a weight update. The harness built for GPT-OSS-120B fixes 91.8\% of the SEC-bench CVEs in a single attempt, on par with the 87.0\% and 92.0\% that EvoRepair and ContraFix reported for the proprietary GPT-5-mini with several attempts per vulnerability~\cite{evorepair,contrafix}.

\textbf{Security of the build loop.} \rb{} turns training traces into tools and instructions, so it inherits the risks of any system that learns from untrusted data. First, the code that the agent reads may contain instructions aimed at the model, for example a comment that asks it to delete a check. Such prompt injection affects any repair agent, and the built harnesses add no new channel for it. Second, an attacker who controls training vulnerabilities could try to steer the optimizer toward a harmful tool or instruction. The gate limits this risk, because an edit must improve the success rate on held-out tasks, and the built harness remains a short, readable file that a defender can audit before deployment. Third, the agent may learn to satisfy the oracle without fixing the vulnerability. The training oracle of \rb{} rejects patches that only mask the PoV, and the gains carry over to the harder CyberGym-E2E and PatchBench. Finally, like any method that improves security agents, \rb{} is dual-use. In this work, we apply it only to vulnerability repair.

\textbf{Toward self-improvement.} The optimizer of \rb{} is itself an agent with a harness, so the same loop could in principle improve the optimizer. Our results indicate where this is feasible. A model gains little from optimizing its own harness (RQ5), but a stronger model builds an effective harness for a weaker one, and harnesses built by one model can raise another (RQ6). One practical path is to let each generation of models build the harnesses of the next.

\section{Related Work}
\label{sec:related}

\textbf{Automated vulnerability repair.} Early learning-based approaches translated vulnerable code into fixed code~\cite{vrepair,vulrepair}, and program-analysis approaches derived a fix from the crash condition~\cite{extractfix}. LLMs have been shown to repair many vulnerabilities from a prompt alone~\cite{pearce}. Recent systems, some built for the DARPA AI Cyber Challenge~\cite{aixcc}, have used agents that iterate with the compiler and the crashing input~\cite{patchagent,san2patch,googlepatching,pagent}. General software engineering agents have followed the same pattern~\cite{sweagent,openhands,agentless}. These harnesses were designed by hand for the models available to their authors. \rb{} builds the harness automatically for each model.

\textbf{Harness building for vulnerability repair.} EvoRepair~\cite{evorepair} and ContraFix~\cite{contrafix} improve a repair agent from its own repair episodes and are the state of the art on SEC-bench. EvoRepair extracts experiences from repair trajectories, scores them by quality and retrieves them to guide later repairs. ContraFix compares the runtime states of crashing and non-crashing variants of the PoV to locate the repair, and it stores validated repair episodes as mutation and correction skills. Both focus on learning the content given to an agent whose architecture stays fixed, and neither evaluates a new entry on held-out tasks before keeping it. Memories of past repairs~\cite{memrepair,cyberevolver}, libraries of skills, insights or workflows~\cite{voyager,expel,awm} and verbal feedback within a task~\cite{reflexion,selfrefine} share this design. \rb{} also changes the tools, agents and runtime controls of the harness, and it keeps an edit only if a statistical gate on held-out tasks supports it (Table~\ref{tab:related}).

\textbf{Prompt and program optimization.} Several methods optimize the text given to an LLM, using model-generated candidates~\cite{opro,promptbreeder}, compiled pipelines~\cite{dspy}, textual feedback that acts as a gradient~\cite{textgrad,gepa} or an evolving context~\cite{ace}. SelfOp optimizes a security agent by applying batched textual feedback to a single skill file~\cite{selfop}. These methods focus on editing the text of a fixed pipeline, whereas \rb{} can also add tools and agents and change how they delegate.

\begin{table}[t]
    \caption{How \rb{} differs from related approaches ($\sim$: in some of the methods).}
    \label{tab:related}
    \centering\footnotesize
    \setlength{\tabcolsep}{2.4pt}
    \newcommand{\y}{\ding{51}}\newcommand{\n}{--}
    \resizebox{\columnwidth}{!}{\begin{tabular}{@{}lccccc@{}}
    \toprule
    & \multicolumn{1}{p{1.05cm}}{\centering Edits structure} & \multicolumn{1}{p{1.15cm}}{\centering Reads failed traces} & \multicolumn{1}{p{1.05cm}}{\centering Held-out gate} & \multicolumn{1}{p{1.05cm}}{\centering Weights fixed} & \multicolumn{1}{p{1.05cm}}{\centering Vuln.\ repair} \\
    \midrule
    EvoRepair, ContraFix~\cite{evorepair,contrafix} & \n & \y & \n & \y & \y \\
    Prompt optimization~\cite{textgrad,gepa,dspy} & \n & \y & \y & \y & \n \\
    Agent design search~\cite{adas,aflow,dgm} & \y & $\sim$ & \y & \y & \n \\
    Meta-Harness~\cite{metaharness} & \y & \y & $\sim$ & \y & \n \\
    RLVR~\cite{grpo,deepseekr1} & \n & \n & \n & \n & \n \\
    \rb{} & \y & \y & \y & \y & \y \\
    \bottomrule
    \end{tabular}}
\end{table}

\textbf{Automated design of agentic systems.} Other work searches over agent programs and workflows~\cite{adas,aflow,maas}, including multi-agent harnesses for vulnerability discovery~\cite{agentflow}, or lets an agent rewrite its own code~\cite{stop,dgm}. Meta-Harness~\cite{metaharness} lets a coding agent rewrite the harness code of an LLM system from the scores and traces of earlier candidates. Most of this work has targeted question answering, mathematics or code generation, where a candidate design is scored cheaply and the search can try many candidates. In vulnerability repair, every evaluation runs agents in containers and rebuilds the project. \rb{} therefore proposes few candidates per round, grounds each in a diagnosis of failed and successful episodes and accepts it only after a paired statistical test with a confirmation on fresh runs.

\textbf{Weight updates.} RLVR updates the model weights from verifiable rewards~\cite{grpo,deepseekr1,swerl}, often through low-rank adapters~\cite{lora}. Model editing changes a small set of weights to alter a specific behavior~\cite{rome}, and WSP lets an LLM write such edits~\cite{wsp}. Both need access to the weights. \rb{} leaves the weights untouched and needs no GPU beyond inference, and our comparison shows that it learns faster than both from the same data.

\section{Conclusion}

We presented Reinforcement Building (\rb{}), a framework that automatically builds the harness that a given model uses to repair vulnerabilities, without updating its weights. In each round, an optimizer reads the successful and failed episodes of the model and proposes edits to the harness, and a statistical gate keeps only the edits that improve it on held-out tasks. Because the optimizer reads the traces of the failed episodes, \rb{} can learn even when the model almost never succeeds, without a reward designed in advance for its failures. With 190 training episodes, \rb{} raised the SEC-bench success rate of Gemma-4-31B by 71\%, from 45.5\% to 77.7\%, above the state-of-the-art harness builders for vulnerability repair and the weight updates trained on the same data. It raised that of Gemma-4-E4B by a factor of 8.5, from 4.8\% to 40.8\%, where RLVR was not able to obtain significant downstream gains. On GPT-OSS-120B, a model of another family, it raised the SEC-bench success rate from 59.5\% to 91.8\%. With Muse-Glimmer-30B as the optimizer, \rb{} still improved Gemma-4-31B by 64\% to 100\% relative to the base harness on the three benchmarks. As new models are released, \rb{} can build a specialized defensive harness for each of them from a few hundred training episodes.

\bibliographystyle{IEEEtran}
\bibliography{refs}

\clearpage
\raggedbottom  
\appendices

\section{Base Harness}
\label{app:vanilla}
The base harness $\Theta_0$ is a single agent with a shell, a budget of 150 steps and the repair instructions of SEC-bench, which we quote verbatim. The built-in commands \texttt{rb-build} and \texttt{rb-repro} build the project and run the PoV.
\begin{promptbox}
\begin{verbatim}
You are a software engineer fixing a security
vulnerability.
Your task is to make the minimal changes to non-test
files in the repository to ensure the crash points
specified in the issue description are not triggered.

Follow these steps to resolve the issue:
1. EXPLORATION: explore the repository structure using
   tools like `find` and `grep`; identify the files
   mentioned in the bug description and locate where
   the vulnerability exists.
2. ANALYSIS: think carefully about the vulnerability
   and propose 2-3 possible approaches to fix it;
   analyse the root cause and select the most promising
   approach.
3. IMPLEMENTATION: edit the source code to implement
   your chosen solution. Make minimal, focused changes
   and do not introduce new security issues.
4. VERIFICATION: run `rb-build` to build the project;
   if it succeeds, run `rb-repro` to verify the fix
   prevents the crash. If the fix fails, revise and
   repeat until `rb-repro` shows that the sanitizer
   error is NOT triggered. This is the success
   criterion.
5. FINAL REVIEW: re-read the bug description, review
   your diff, and make sure the fix is minimal.
\end{verbatim}
\end{promptbox}

\section{Harness Built for Gemma-4-31B}
\label{app:h31b}
The harness that \rb{} built for Gemma-4-31B in four rounds keeps a single agent. It adds a step-level loop guard (\texttt{loop\_guard: 3}), two restarts with a note, the tool in Table~\ref{tab:tools31b} and the following instructions, quoted verbatim.
\begin{promptbox}
\begin{verbatim}
You are a software engineer fixing a security
vulnerability (a sanitizer crash) in this C/C++
repository. Make the minimal change to non-test source
files so that the crash described in the issue no
longer happens.

Think in comments: start EVERY command block with one
or two `# ` lines: what the last output told you and
what you check next and why (your current hypothesis,
not a restatement). They cost no step. Never send a
command without them.

1. Run `rb-repro` once to see the crash stack. Open the
   project's own frames (skip sanitizer/fuzzer-runtime
   frames) with `sed -n 'A,Bp' FILE`. Read each region
   ONCE: never re-run a command you already ran, its
   output is still above.
2. Within about 15 commands write `# ROOT CAUSE: ...`
   and `# FIX: ...` comments and act on them. Decide
   the root cause: which length, index, pointer or
   state is invalid at the crash, and where it should
   have been checked. Prefer adding the missing check
   (bounds/length/NULL/overflow/state) with the
   function's existing error-handling convention, at
   the place where the bad value is used or enters; do
   not just delete the crashing statement.
3. Edit with the `replace` tool (copy the old lines
   exactly from the file; it prints the edited lines):
   replace FILE <<'END'
   <<<<<<< SEARCH
   old lines
   =======
   new lines
   >>>>>>> REPLACE
   END
   For a one-line change `sed -i 'Ns/old/new/' FILE` is
   fine too.
4. Verify: `rb-build >/tmp/build.log 2>&1; echo
   build_exit=$?; tail -20 /tmp/build.log` then
   `rb-repro`. If the build fails, fix the compile
   error. If it still crashes, read the NEW top frame
   and adjust the fix.
5. When rb-repro shows no sanitizer report, run `git
   diff` once and finish (reply without a command) with
   a short summary.

Getting unstuck: if you get a repeated-command warning,
notice you are re-reading code you already saw, or have
used about 45 commands without a clean rb-repro, call
the `restart` tool with a note (under 1500 characters;
it is a tool call like a function, not a bash command):
crash location (file:line, function), your root-cause
hypothesis, what you changed and why it failed, what to
try next. A restart resets the tree and gives you a
fresh context; it beats looping. On a restarted
attempt, start from the note: go straight to the
suspected code and edit. Never restart while your patch
builds and rb-repro is clean.
\end{verbatim}
\end{promptbox}

\begin{table}[!htbp]
\caption{Tool of the harness built for Gemma-4-31B, with its length in lines of Python and the description that the agent sees.}
\label{tab:tools31b}
\centering\scriptsize
\begin{tabular}{@{}lrL{5.6cm}@{}}
\toprule
Tool & Lines & Description \\
\midrule
\texttt{replace} & 181 & Exact block edit of a source file. Usage: \texttt{replace FILE} with stdin holding one or more blocks \texttt{'<{}<{}<{}<{}<{}<{}< SEARCH\textbackslash{}n<old lines copied from the file>\textbackslash{}n=======\textbackslash{}n<new lines>\textbackslash{}n>{}>{}>{}>{}>{}>{}> REPLACE'}. From bash: \texttt{replace FILE <{}<'END' ... END}. Falls back to whitespace-insensitive line matching (keeps the file's indentation); refuses ambiguous or missing matches without changing anything; prints the edited lines with numbers. \\
\bottomrule
\end{tabular}
\end{table}

\section{Harness Built for Gemma-4-E4B}
\label{app:he4b}
The harness that \rb{} built for Gemma-4-E4B in eleven rounds also keeps a single agent, with the loop guard, three tools (Table~\ref{tab:toolse4b}) and two memory notes. Its instructions and notes are quoted verbatim.
\begin{promptbox}
\begin{verbatim}
You are a software engineer fixing a security
vulnerability (a sanitizer crash) in a C/C++ project.
Make a minimal change to non-test source files so the
PoC no longer triggers the sanitizer report, without
breaking valid inputs.

Tools (run them from bash):
- `showsrc FILE LINE` prints numbered lines around LINE
   (`showsrc FILE START-END` for a range). FILE may be
   the path exactly as printed in the stack trace; the
   tool finds it in the repo. Use it instead of cat/sed
   -n.
- `editsrc FILE START END <<'EOF'` ...new lines...
   `EOF` replaces lines START..END (inclusive) with the
   new lines and prints the result. Use it instead of
   `sed -i`/perl (sed silently changes nothing when its
   pattern does not match, and breaks on [ ] * / &).
   `showsrc` the exact lines just before editing so the
   line numbers are current; to insert code, repeat the
   original line in the new text.
- `fixcheck` rebuilds the project and runs the PoC. It
   prints "result: build failed" (with the compiler
   errors), "result: still crashes" (with the sanitizer
   stack) or "result: fixed locally". `rb-repro` alone
   does not rebuild, so after every edit use
   `fixcheck`.

Steps:
1. Read the stack trace in the issue. `showsrc` the
   crashing line and the callers in the project's own
   code (top 3-4 frames). Work out which size, index,
   length or pointer is invalid and where it comes
   from.
2. Fix the root cause: add the missing
   bounds/length/null-pointer/overflow check, or
   correct the wrong size computation, where the bad
   value is produced or first used. Fail the same way
   the surrounding code reports errors (return its
   error code, false, a null pointer, ...). Do not edit
   fuzzer harnesses or tests, and do not remove
   functionality.
3. Run `fixcheck`. If build failed: `showsrc` the
   reported line and repair it. If still crashes: read
   the new stack; if it is the same line, your check is
   not on the executed path, so look again at the
   values involved.
4. Never finish unless the latest `fixcheck` printed
   "result: fixed locally". You have about 150 steps;
   stopping early with a crashing, non-building or
   empty patch is a failure. If one approach fails 3
   times, undo it with `git checkout -- FILE` and try a
   different place or approach.
5. After the fix works, run `git diff` once, make sure
   it contains only the intended fix, then finish.
Each reply: brief reasoning, then exactly one
```bash``` block with one command.
\end{verbatim}
\end{promptbox}

\textbf{Memory notes.}
\begin{promptbox}
\begin{verbatim}
- Reading the sanitizer report: buffer overflow
   (heap/stack/global) -> an index or length used at
   the crashing line exceeds the buffer; find where
   that value comes from (usually parsed input) and
   check it against the real buffer size before it is
   used. use-after-free / double-free -> compare the
   'freed by' stack with the crashing stack: the object
   is freed while still referenced; clear the stale
   pointer after freeing, or avoid the free/use on that
   path. segfault on a null or small address -> a
   pointer from a failed allocation, lookup or parse is
   used without a check. memory leak -> free what was
   allocated on the early-return/error path. runtime
   error (integer overflow, bad shift, misaligned or
   out-of-range value) -> validate the operands before
   the operation. Fix the project's code, never the
   fuzz harness, as close as possible to where the bad
   value enters.
\end{verbatim}
\begin{verbatim}
- When fixcheck prints 'result: build failed', assume
   your edit caused it and never finish in that state.
   Read the first error, the diff and the source lines
   fixcheck prints. Use only names that already exist
   in that code (variables, struct members, macros,
   error codes, labels; check them with grep -n)
   instead of guessing new ones. If repeated edits left
   the file tangled, restore it with the git checkout
   command fixcheck prints and redo the fix as one
   small, complete edit.
\end{verbatim}
\end{promptbox}

\begin{table}[!htbp]
\caption{Tools of the harness built for Gemma-4-E4B.}
\label{tab:toolse4b}
\centering\scriptsize
\begin{tabular}{@{}lrL{5.6cm}@{}}
\toprule
Tool & Lines & Description \\
\midrule
\texttt{fixcheck} & 384 & Rebuild the project (rb-build) and, if the build succeeds, run the PoC (rb-repro). Prints a short result: build failed + first compiler errors, the current source at the first error, your current git diff, brace-balance warnings, similar existing names for unknown identifiers/members and the exact restore command; still crashes + sanitizer stack; or fixed locally. No arguments. Run it after every edit. \\
\texttt{showsrc} & 115 & \texttt{showsrc FILE LINE [CONTEXT]} | \texttt{showsrc FILE START-END} | \texttt{showsrc FILE:LINE} -- print numbered source lines. FILE may be a path from a stack trace (it is located in the repo by suffix). \\
\texttt{editsrc} & 227 & \texttt{editsrc FILE START END <{}<'EOF'} (new lines) \texttt{EOF} -- replace lines START..END (inclusive) of FILE with the text on stdin, then print the edited region (warns if the brace balance changed). \texttt{editsrc FILE START END delete} deletes the lines. Use showsrc first to get current line numbers. \\
\bottomrule
\end{tabular}
\end{table}

\section{Typed Edits}
\label{app:edits}
Table~\ref{tab:edits} lists the edits that a plan may contain. The runtime applies them in the order deletion, modification, creation, delegation and entry point, and rejects a plan whose edits do not apply.

\begin{table}[!htbp]
\caption{Typed edits of the harness.}
\label{tab:edits}
\centering\footnotesize
\begin{tabular}{@{}lL{4.6cm}@{}}
\toprule
Edit & Effect \\
\midrule
\texttt{add/edit/remove\_agent} & create, rewrite or delete an agent: instructions, tools, step budget, loop guard \\
\texttt{add/edit/remove\_tool} & create, rewrite or delete a tool: description and script \\
\texttt{add/remove\_note} & add or delete a memory note shown to every agent \\
\texttt{grant/revoke} & add or remove a delegation edge between two agents \\
\texttt{set\_entry} & choose the agent that receives the task \\
\bottomrule
\end{tabular}
\end{table}

\end{document}